# Magnetic domain structure and domain wall bound states of topological semimetal EuB$_6$

Qiang Li[1,2]*, Haiyang Ma[3,4]*, Xin Zhang[3], Yanfeng Guo[3,4], Jianpeng Liu[3,4], Wenbo Wang[3,4]†, Xiaodong Zhou[1,7,8]† and Jian Shen[1,2,5,6,7,8]†

[1]State Key Laboratory of Surface Physics and Institute for Nanoelectronic Devices and Quantum Computing, Fudan University, Shanghai 200433, China.

[2]Department of Physics, Fudan University, Shanghai 200433, China.

[3]School of Physical Science and Technology, ShanghaiTech University, Shanghai 201210, China

[4]ShanghaiTech Laboratory for Topological Physics, ShanghaiTech University, Shanghai 201210, China

[5]Shanghai Research Center for Quantum Sciences, Shanghai 201315, China

[6]Collaborative Innovation Center of Advanced Microstructures, Nanjing 210093, China

[7]Shanghai Qi Zhi Institute, Shanghai 200232, China

[8]Zhangjiang Fudan International Innovation Center, Fudan University, Shanghai 201210, China

*These authors contributed equally to this work
†Emails:
wangwb1@shanghaitech.edu.cn,
zhouxd@fudan.edu.cn,
shenj5494@fudan.edu.cn

**Abstract:**

Electrically manipulating domain wall (DW) is at central in the field of spintronics. Magnetic Weyl semimetal (WSM) offers an additional knob for electric manipulation of DW by utilizing the topological DW bound states. Using magnetic WSM material EuB$_6$ as a prototype system, we investigate its DW bound states combining first principles calculations and domain imaging via magnetic force microscopy (MFM). The MFM measurements reveal that domains with magnetization aligned along three [100] directions dominate over others under a small magnetic field along [001] direction. Accordingly, first principles calculations are performed based on experimentally determined DW type, which show a robust DW bound state featuring a spectral distribution over a large portion of the DW Brillouin zone. Such DW bound states should result in significant amount of localized charges which can be used to drive DW motion via the electrostatic forces exerted on the localized DW charges.

**Introduction**

Controlling magnetic domain wall (DW) motion is important for spintronic applications including memory and logic devices [1,2]. In general, DW motion can be achieved via spin-transfer torque [3-5] or spin-orbit torque [6-8]. The latter is effective for materials with strong spin-orbit-coupling (SOC) including topological materials which possess strong SOC and spin-momentum locking [9]. Recent theoretical study suggests that DWs in magnetic Weyl semimetal (WSM) host unique topological bound states which can in turn affect the spin texture (chirality) or the motion of DWs under an electric field in a manner similar to spin-transfer torque [10-13]. However, these arguments are rather generalized and the knowledge of topological DW bound state in a real magnetic WSM is very limited, impeding the experimental exploration of such new DW control methods.

Soft ferromagnetic (FM) material $EuB_6$ has long been studied for which the concept of magnetic polarons was discussed as the underlying mechanism for its large magnetoresistance [14-23]. Research interests are revived as a recent theoretical calculation suggests that $EuB_6$ can be put into different topological semimetal phases by simply changing the magnetization direction in its FM state [24]. For instance, $EuB_6$ is a topological nodal-line semimetal when the moment is aligned along the [001] direction, and it evolves into a WSM when the moment is rotated to the [111] direction. Coexistence of a nodal line and Weyl points is found with the moment in the [110] direction. Angle-resolved photoemission spectroscopy measurements on $EuB_6$ single crystal have revealed that the band splitting and inversion occur upon FM transition confirming such time-reversal-symmetry breaking induced topological phases [25,26]. The magnetotransport measurements show weak magneto-crystalline anisotropy consistent with the soft FM state of $EuB_6$ [27].

The close correlation between magnetization direction and topological phase in EuB$_6$ implies that its magnetic DW, i.e., crossover region between domains with different magnetization directions and thus different topological phases, may host a topological bound state. Here we perform first principles calculations on such topological DW bound states in light of their potential applications in DW control. These calculations are based on the DW types in EuB$_6$ determined by our cryogenic magnetic force microscopy (MFM) imaging. In particular, we use MFM to investigate the magnetic domain structure on the (001) surface of EuB$_6$ single crystal. At zero magnetic field, one observes the formation of large in-plane domains on the surface with typical size of tens of $\mu m$. A small magnetic field along [001] direction breaks up these large in-plane domains into a mixture of small in-plane and out-of-plane domains with magnetization orientation along [100], [010] and [001] directions. Guided by such experimental observation, our first principles calculations focus on DWs separating domains of [100], [010] and [010] directions, which confirm the existence of robust topological bound states in DWs.

## Results

**Global magnetization characterizations**

EuB$_6$ crystallizes in a simple cubic lattice similar to other metal hexaboride $R$B$_6$ ($R$= rare earth or alkaline metals, see inset of Fig. 1(a)). We have done x-ray diffraction to confirm the single crystallinity of EuB$_6$ sample [28,29] (see Supplemental Material [30]). Figure 1(a) shows the temperature-dependent resistivity of EuB$_6$ single crystal. A resistivity upturn is observed at 15 K corresponding to an FM transition as seen from the temperature-dependent magnetization (inset of Fig. 1(b)) [14,15]. For a high symmetric cubic crystal, the magneto-crystalline anisotropy energy can be phenomenologically expressed by two constants $K_1$ and $K_2$: $E_k = K_1(S_x^2 S_y^2 + S_y^2 S_z^2 + S_z^2 S_x^2) + K_2 S_x^2 S_y^2 S_z^2$. They together determine the easy axis of an FM order. Early experiment suggests a two-step FM transition in EuB$_6$ with an easy axis along [100] below $T_{c1} = 15.3$ K and along [111] below $T_{c2} = 12.5$ K [15,20,31]. We examine the magnetic anisotropy of our EuB$_6$ single crystal by comparing the bulk initial magnetization curves along three representative directions, i.e., [100], [110] and [111] directions. It appears that [100] is the easy axis of the FM ground state because the magnetization along [100] rises up faster and saturates earlier than the other two directions (Fig. 1(b)). Accordingly, [111] is the hard axis. One could further estimate the magnetic anisotropy from the initial magnetization curves by considering the energy differences needed for saturation along different axes [32]. The obtained anisotropy energy differences are 0.022 meV between [110] and [100] and 0.039 meV between [111] and [100], which are small indicating that EuB$_6$ is a very soft FM material. Note that the saturation magnetization is close to 6.5$\mu_B$ per Eu ion at 2 K suggesting a fully polarized spin state of Eu $4f$ electrons.

**MFM characterizations**

The EuB$_6$ single crystal was cleaved in air to expose the (001) surface. After the cleavage, we use both optical microscope and atomic force microscopy to check the cleaved surface at room temperature in air to ensure its high quality (see Supplemental Material [30]). MFM measurements were performed at 6 K on the (001) surface of EuB$_6$ single crystal to characterize the magnetic domain structure in its FM ground state. Although the three-dimensional (3D) domain structure is inaccessible by probes with only surface sensitivity, valuable insights can be obtained from MFM measurements with aid of domain theory. Figure 2(a) is an MFM image consisting of multiple scans stitched together to cover a large (001) surface area. This image was taken at zero magnetic field after a zero-field cooling. Except for some topographic features such as terraces and step edges (see Supplemental Material [30]), Fig. 2(a) is largely featureless over the whole area. The most eye-catching features are several dark lines corresponding to DWs. While some of these DWs align along [110] or [1-10] directions (denoted by red arrows in Fig. 2(a)), others meander on the surface (denoted by black arrows in Fig. 2(a)). The featureless morphology of Fig. 2(a) reflects that the domains are all in-plane magnetized due to the large stray field energy, which are not detectable in MFM as it only senses the out-of-plane stray field. Only at the DWs can one get out-of-plane magnetic moments due to the winding of magnetization across the DWs. This gives rise to the dark contrast in MFM images.

The in-plane magnetization direction of each domain in Fig. 2(a) cannot be directly identified by MFM. However, with the aid of domain theory, important insights are obtained. In Fig. 2(b), we show an MFM image of the same area as Fig. 2(a) with a small applied field (500 Oe) along [001] direction, i.e., surface normal direction. The domain pattern changes dramatically from the zero field case. Domain with sawtooth shape grows from the dark DW lines, which is known as fir tree pattern commonly observed on the surface of soft magnetic material [32,33]. We draw a 3D schematic in Fig. 3(m) to illustrate the magnetization orientation of a typical fir tree domain pattern. It can be considered as the newly developed fir tree 90° domains (bright yellow) superimposed on the original basic 180° domains (grey yellow). The principle underlying the observed fir tree pattern is the introduction of shallow surface domains collecting the net flux for the case of slightly misoriented surfaces, i.e., the [100] easy axis cants slightly from in-plane to out-of-plane under the 500 Oe field along [001] direction. This canting leads to the misorientation between the magnetization and the surface and explains the observed domain contrast in Fig. 2(b) compared to zero field case of Fig. 2(a). The more it cants, the darker it appears in Fig. 2(b). Another important insight from the observation of the fir tree pattern is that DWs at zero field in Fig. 2(a) are mostly 180° ones, so the orientation of the DWs largely reflects the direction of in-plane magnetization of the neighboring domains. The meandering DWs of Fig. 2(a) thus suggest that the in-plane magnetization is almost free to rotate on the surface of $EuB_6$, which is reasonable because the thermal energy at the measuring temperature (6 K) is larger than the anisotropy energy difference between different in-plane directions. We note that, the experimentally observed fir tree pattern in Fig. 2(b) is richer than what is shown in the schematic of Fig. 3(m), i.e., in additional to the two jagged sides as shown in the schematic, Fig. 2(b) also features one jagged and one straight side. The fir tree pattern has different variants depending on the relative misorientation between the surface the nearest magnetization easy axis [32]. Given that in-plane magnetization easy axis is almost free to rotate on the surface, it is not surprising to find a rich fir tree pattern under the field.

Figure 3(a)-(l) show evolution of the domain pattern in a chosen area (red square in Fig. 2(b)) with further increasing field over 500 Oe along [001] direction. The fir tree patterned domains break into smaller ones (Fig. 3(a) to (d)), and gradually form a palm tree like pattern whose branches and leaves are vividly displayed (Fig. 3(e) to (h)). As schematically shown in Fig. 3(n), the alternating horizontal dark and grey yellow domains in Fig. 3(h) correspond to the out-of-plane and in-plane domains with magnetizations along [001] and [100] ([-100]) directions, respectively. The in-plane domains form additional head-to-head and tail-to-tail DWs corresponding to the vertical dark and bright yellow lines in Fig. 3(h). These DWs have finite out-of-plane magnetic moments due to the closure of magnetic flux giving rise to bigger MFM signal contrast. Figure 3(o) shows the cross-sectional view of the marked yellow regions in Fig. 3(n). Inside the sample, there are domains with magnetization along [001] direction. However, the magnetic flux has to be closed on the surface to minimize the stray field energy. Therefore, one has in-plane domains on the top surface whose magnetization forms a close loop with the internal domains, i.e., V-lines based on domain theory [32]. The fractal nature of the palm tree domain pattern is attributed to a so-called domain branching process [32]. Such domain branching is aimed to completely close the magnetic flux by continually breaking up large domains into small pieces. That explains why the out-of-plane domains in Fig. 3(h) are separated by so many small in-plane domains in between (see Fig. 3(n)). This fragmentation of out-of-plane domains effectively reduces the stray field energy. It becomes prominent for materials with weak magnetic anisotropy that allows the magnetization to almost freely rotate without much energy cost. Such domain branching is also expected for EuB$_6$ which has a relatively large magnetization ($M_s \sim 6.5 \mu_B$/Eu) and tends to form more magnetic domains on sample surface to minimize stray field energy, which is proportional to $(M_s)^2$. When the field continues to increase (Fig. 3(i) to (l)), dark out-of-plane domains expand at the cost of in-plane domains. Note that only horizontally oriented [100] domains are observed in Fig. 3(j) and (k), which is due to the limited size of the displayed area rather than any C4 symmetry breaking cause. We show areas in the supplementary (see Supplemental Material [30]) displaying both [100] and [010] domains under this field range. The system eventually becomes uniformly magnetized at 6800 Oe which agrees with the global $M(H)$ behavior in Fig. 1(b). Note that the feature seen in Fig. 3(l) is the topographic terrace in this area.

**DW bound states calculations**

Having known the domain structure and the DW type, we proceed to perform first principles calculations on the corresponding DW bound states. These DW bound states can be considered as two-dimensional (2D) topological interface states. To begin with, the bulk band structures of $EuB_6$ in an FM state with different magnetization directions have been calculated, which are consistent with the previous results confirming $EuB_6$ as a topological semimetal [24] (see Supplemental Materials [30]). After that, as a comparison, we first calculate the regular topological surface states (interfaced with vacuum) of $EuB_6$. The (100) surface band structure of $EuB_6$ with magnetization along [100] direction is presented in Fig. 4(a), where surface states are only visible around the $\Gamma$ point. In Fig. 4(b) we further show the surface spectral function at Fermi level (set to zero) in the (100)-surface Brillouin zone (BZ), from which it is more clearly seen that the spectral weights are mainly located at the central $\Gamma$ point and $X/Y$ points. They are the surface drumhead states strictly confined to the interior of the surface projection of the nodal loop, forming the red-disk shaped surface state as shown in Fig. 4(b). We then construct a tail-to-tail DW (see Supplemental Materials [30] for head-to-head DW results) separating domains with magnetizations along [-100] and [100] directions, as denoted in the inset of Fig. 4(c). The electronic bound state of such 180° DW is shown in Fig. 4(c) with the band dispersions plotted along a high-symmetry path within the (100) DW BZ. Different from the surface states shown in Fig. 4(a), we see that the DW bound states expand over most of the DW BZ with an energy dispersion ~ 0.4 eV. In Fig. 4(d) we further show the DW spectral function at Fermi level throughout the DW BZ, from which the two Fermi surfaces contributed by DW bound states are evidently seen. We also calculate 90° DWs separating domains along [100] and [001] directions as another experimentally observed DWs (Fig. 4(e) and (f)). Interestingly, a very similar topological electronic structure bound to the DW is found. As the magnetization is rotated in real space from [100] to [001] direction from one domain to the other, the Weyl nodal loops in reciprocal space are also rotated by 90°, leaving a trace of the nodal points in the DW BZ which is filled by the topological DW bound states (Fig. 4(g)). That explains why DW bound state has a wider spectral distribution on DW BZ than that of regular surface states.

## Discussions

Our global magnetization and microscopic MFM domain characterizations reveal $EuB_6$ as a soft FM material with a small magnetic anisotropy and a large saturation magnetization (spin polarized state). They dictate the domain structure at zero field (Fig. 2(a)), i.e., domains with in-plane magnetization are formed on the surface to minimize the stray field energy which is proportional to $(M_s)^2$. In addition, the orientation of magnetization is free to rotate on the surface as a result of small magnetic anisotropy. They also manifest in the domain evolution, i.e., a domain branching process proceeds with increasing field. The 3D nature of domain structure in $EuB_6$ can be unveiled by the observation of V-lines. Moreover, the fact that an applied field along [001] direction drives the system into a domain structure with domains mostly along [100] and [010] directions (Fig. 3(e) to (h)) on the surface indicates a 3D Cubic symmetry rather than a uniaxial symmetry.

Our first principles calculations show the robust topological DW bound states resulting from the strong coupling between nodal points (loops) in reciprocal space with the magnetizations in real space, such that they are transformed to different valleys of BZ in different magnetic domains, leaving gapless DW states filling up the interstitial regions. Such internal interface states are different from the conventional topological surface states of WSM which are usually confined to a small region on the surface BZ [34-36]. Since the topological DW bound states occupy most regions of the DW BZ, significant amount of localized charges (on the order of one per primitive cell) can be accumulated to the DW. As a result, an external electric field perpendicular to the DW may drive the DW motion due to the electrostatic forces exerted on the localized DW charges. This is particularly important for the $EuB_6$ system because conventional current-driven DW motion may be hindered due to the fact that the Fermi surfaces of two different magnetic domains of $EuB_6$ have little overlaps in momentum space, which is difficult to transmit charge current through the DW from a ballistic transport perspective. The electrostatic forces experienced by the topological DW charges may be the dominant driving forces for the DW motion, which is also desirable from energy saving point of view.

## Summary

In conclusion, we systematically studied the magnetic domain structures of an FM order on the (001) surface of $EuB_6$ single crystal at various magnetic fields. This microscopic characterization confirms $EuB_6$ as a 3D soft FM material with a small magnetic anisotropy and a large saturation magnetization. The calculated topological DW bound states in $EuB_6$ turn out to occupy most regions of the DW BZ, which are qualitatively different from conventional surface states of WSM. While the conventional current-driven DW motion through the spin-transfer torque may be difficult in this system, we argue that the electrostatic forces exerted on the localized charges in the DW could be utilized for the DW motion control. To achieve this goal, one promising route is to synthesize $EuB_6$ 2D thin film as previously demonstrated in $SmB_6$ [37]. The perpendicular magnetic anisotropy may affect the domain structure of $EuB_6$ 2D thin film resulting in a simple domain structure populated with [001] domains, which is convenient for experimental manipulations.

## Acknowledgements

The work at Fudan University is supported by National Natural Science Foundation of China (Grant Nos. 12074080, 11804052), Shanghai Science and Technology Committee Rising-Star Program (19QA1401000). The work at

ShanghaiTech University is supported by National Natural Science Foundation of China (Grant Nos. 11874262, 12174257), Science and Technology Commission of Shanghai Municipality (Grant No. 21PJ410800), and National Key R & D program of China (Grant No. 2020YFA0309601).

# References

[1]   I. Zutic, J. Fabian, and S. Das Sarma, Rev Mod Phys **76**, 323 (2004).
[2]   S. S. P. Parkin, M. Hayashi, and L. Thomas, Science **320**, 190 (2008).
[3]   G. Tatara, H. Kohno, and J. Shibata, Phys. Rep. **468**, 213 (2008).
[4]   A. Brataas, A. D. Kent, and H. Ohno, Nat. Mater. **11**, 372 (2012).
[5]   T. Okuno, D. H. Kim, S. H. Oh, S. K. Kim, Y. Hirata, T. Nishimura, W. S. Ham, Y. Futakawa, H. Yoshikawa, A. Tsukamoto, Y. Tserkovnyak, Y. Shiota, T. Moriyama, K. J. Kim, K. J. Lee, and T. Ono, Nat. Electron. **2**, 389 (2019).
[6]   I. M. Miron, K. Garello, G. Gaudin, P. J. Zermatten, M. V. Costache, S. Auffret, S. Bandiera, B. Rodmacq, A. Schuhl, and P. Gambardella, Nature **476**, 189 (2011).
[7]   L. Q. Liu, C. F. Pai, Y. Li, H. W. Tseng, D. C. Ralph, and R. A. Buhrman, Science **336**, 555 (2012).
[8]   T. Shiino, S.-H. Oh, P. M. Haney, S.-W. Lee, G. Go, B.-G. Park, and K.-J. Lee, Phys. Rev. Lett. **117**, 087203 (2016).
[9]   D. Pesin and A. H. MacDonald, Nat. Mater. **11**, 409 (2012).
[10]  Y. Araki, A. Yoshida, and K. Nomura, Phys. Rev. B **94**, 115312 (2016).
[11]  A. G. Grushin, J. W. F. Venderbos, A. Vishwanath, and R. Ilan, Phys. Rev. X **6**, 041046 (2016).
[12]  J. P. Liu and L. Balents, Phys. Rev. Lett. **119**, 087202 (2017).
[13]  J. Hannukainen, A. Cortijo, J. H. Bardarson, and Y. Ferreiros, Scipost Phys. **10**, 102 (2021).
[14]  Z. Fisk, D. C. Johnston, B. Cornut, S. von Molnar, S. Oseroff, and R. Calvo, J. Appl. Phys. **50**, 1911 (1979).
[15]  S. Sullow, I. Prasad, M. C. Aronson, J. L. Sarrao, Z. Fisk, D. Hristova, A. H. Lacerda, M. F. Hundley, A. Vigliante, and D. Gibbs, Phys. Rev. B **57**, 5860 (1998).
[16]  P. Nyhus, S. Yoon, M. Kauffman, S. L. Cooper, Z. Fisk, and J. Sarrao, Phys. Rev. B **56**, 2717 (1997).
[17]  P. Das, A. Amyan, J. Brandenburg, J. Müller, P. Xiong, S. von Molnár, and Z. Fisk, Phys. Rev. B **86**, 184425 (2012).
[18]  R. S. Manna, P. Das, M. de Souza, F. Schnelle, M. Lang, J. Müller, S. von Molnár, and Z. Fisk, Phys. Rev. Lett. **113**, 067202 (2014).
[19]  M. Pohlit, S. Rößler, Y. Ohno, H. Ohno, S. von Molnár, Z. Fisk, J. Müller, and S. Wirth, Phys. Rev. Lett. **120**, 257201 (2018).
[20]  D. J. Sivananda, A. Kumar, M. A. Ali, S. S. Banerjee, P. Das, J. Muller, and Z. Fisk, Phys. Rev. Mater. **2**, 113404 (2018).
[21]  S. Rossler, L. Jiao, S. Seiro, P. F. S. Rosa, Z. Fisk, U. K. Rossler, and S. Wirth, Phys. Rev. B **101**, 235421 (2020).

[22] J. Kim, W. Ku, C. C. Lee, D. S. Ellis, B. K. Cho, A. H. Said, Y. Shvyd'ko, and Y. J. Kim, Phys. Rev. B **87**, 155104 (2013).

[23] G. Beaudin, L. M. Fournier, A. D. Bianchi, M. Nicklas, M. Kenzelmann, M. Laver, and W. Witczak-Krempa, Phys. Rev. B **105**, 035104 (2022).

[24] S. M. Nie, Y. Sun, F. B. Prinz, Z. J. Wang, H. M. Weng, Z. Fang, and X. Dai, Phys. Rev. Lett. **124**, 076403 (2020).

[25] S. Y. Gao, S. Xu, H. Li, C. J. Yi, S. M. Nie, Z. C. Rao, H. Wang, Q. X. Hu, X. Z. Chen, W. H. Fan, J. R. Huang, Y. B. Huang, N. Pryds, M. Shi, Z. J. Wang, Y. G. Shi, T. L. Xia, T. Qian, and H. Ding, Phys. Rev. X **11**, 021016 (2021).

[26] W.L.Liu, X.Zhang, S.M. Nie, Z.T. Liu, X.Y.Sun, H.Y.Wang, J.Y.Ding, L.Sun, Z.Huang, H.Su, Y.C.Yang, Z.C.Jiang, X.L.Lu, J.S.Liu, Z.H.Liu, S.L.Zhang, H.M.Weng, Y.F.Guo, Z.J.Wang, D.W.Shen, and Z.Liu, arXiv:2103.04658 (2021).

[27] J. Yuan, X. Shi, H. Su, X. Zhang, X. Wang, N. Yu, Z. Zou, W. Zhao, J. Liu, and Y. Guo, Phys. Rev. B **106**, 054411 (2022).

[28] S. H. Lee, V. Stanev, X. H. Zhang, D. Stasak, J. Flowers, J. S. Higgins, S. Dai, T. Blum, X. Q. Pan, V. M. Yakovenko, J. Paglione, R. L. Greene, V. Galitski, and I. Takeuchi, Nature **570**, 344 (2019).

[29] Z. Wang, W. Han, Q. H. Fan, and Y. M. Zhao, Phys. Status Solidi RRL **15**, 2100249 (2021).

[30] See Supplemental Material at URL for addtional X-ray, MFM and first principles calculations results.

[31] M. L. Brooks, T. Lancaster, S. J. Blundell, W. Hayes, F. L. Pratt, and Z. Fisk, Phys. Rev. B **70**, 020401 (2004).

[32] Alex Hubert and R. Schäfer, *Magnetic Domains: The Analysis of Magnetic Microstructures* (Springer, Berlin, Heidelberg, 1998).

[33] H. J. Williams, R. M. Bozorth, and W. Shockley, Phy. Rev. **75**, 155 (1949).

[34] I. Belopolski, K. Manna, D. S. Sanchez, G. Q. Chang, B. Ernst, J. X. Yin, S. S. Zhang, T. Cochran, N. Shumiya, H. Zheng, B. Singh, G. Bian, D. Multer, M. Litskevich, X. T. Zhou, S. M. Huang, B. K. Wang, T. R. Chang, S. Y. Xu, A. Bansil, C. Felser, H. Lin, and M. Z. Hasan, Science **365**, 1278 (2019).

[35] D. F. Liu, A. J. Liang, E. K. Liu, Q. N. Xu, Y. W. Li, C. Chen, D. Pei, W. J. Shi, S. K. Mo, P. Dudin, T. Kim, C. Cacho, G. Li, Y. Sun, L. X. Yang, Z. K. Liu, S. S. P. Parkin, C. Felser, and Y. L. Chen, Science **365**, 1282 (2019).

[36] N. Morali, R. Batabyal, P. K. Nag, E. K. Liu, Q. A. Xu, Y. Sun, B. H. Yan, C. Felser, N. Avraham, and H. Beidenkopf, Science **365**, 1286 (2019).

[37] Y. F. Li, Q. L. Ma, S. X. Huang, and C. L. Chien, Sci. Adv. **4**, aap8294 (2018).

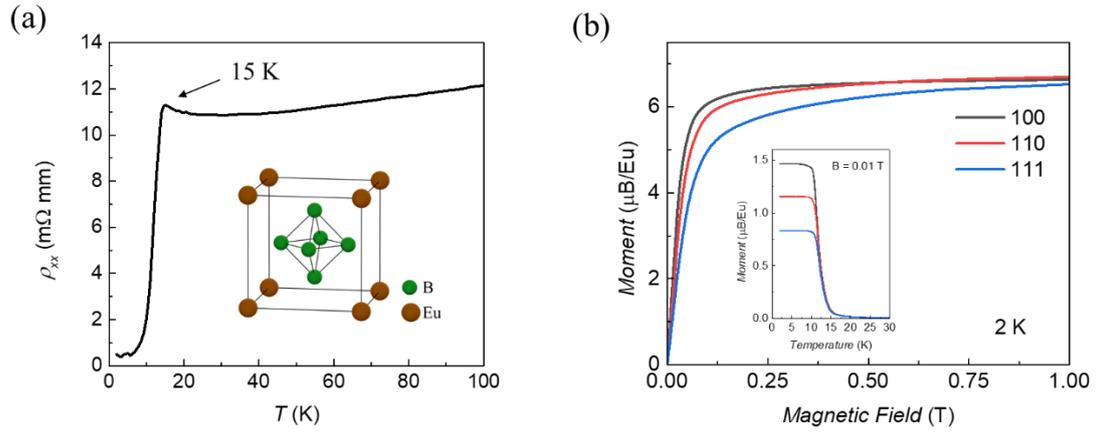

FIG. 1. (a) Temperature-dependent resistivity of EuB$_6$ single crystal. Inset: the crystal structure of EuB$_6$. (b) Initial magnetization as a function of external magnetic field along [100], [110] and [111] directions taken at 2 K. Inset: temperature-dependent magnetization along [100], [110] and [111] directions.

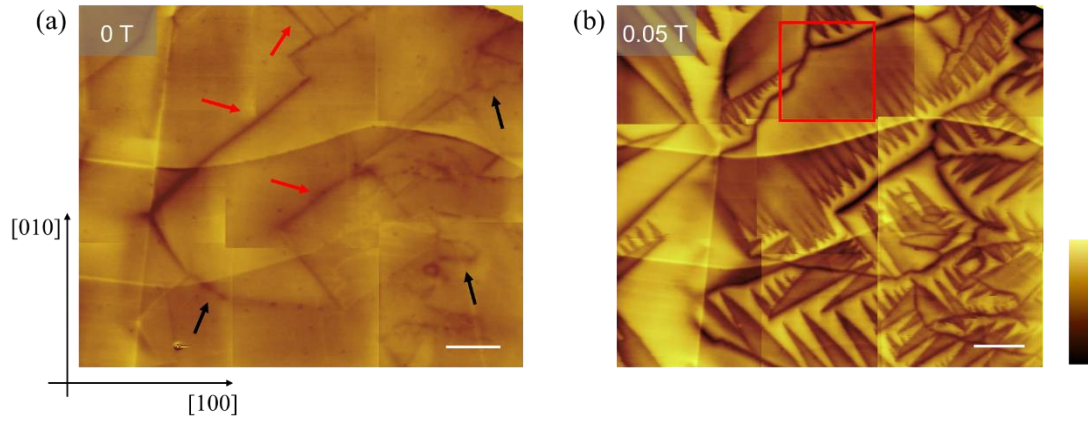

FIG. 2. (a) An MFM image taken at zero field after a zero-field cooling. (b) An MFM image taken under 500 Oe magnetic field along [001] direction. The color scale is 0.6 Hz. The scale bar is 5 $\mu m$.

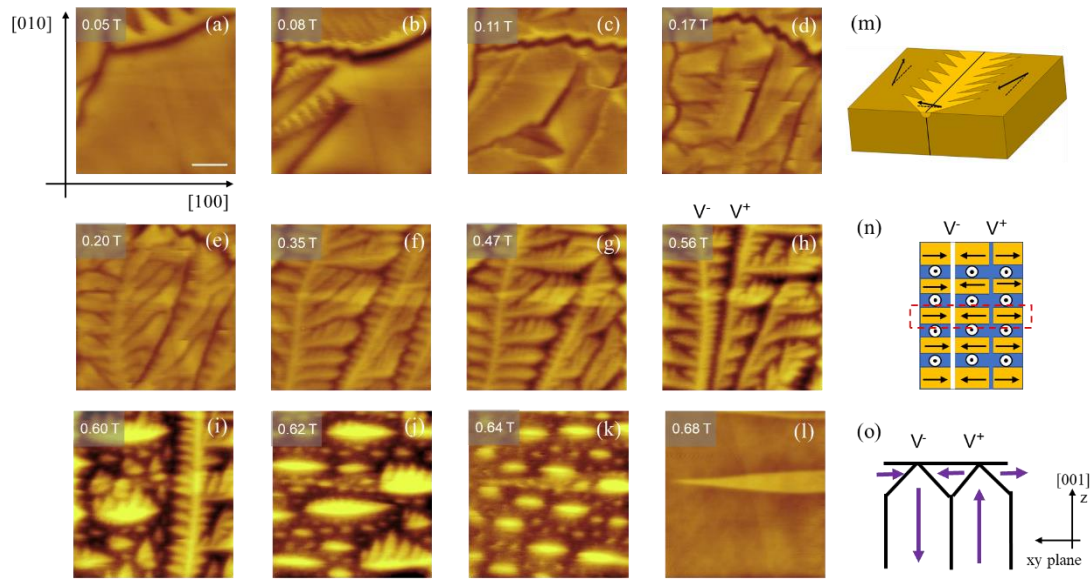

FIG. 3. (a)-(l) Sequential MFM images of the denoted area in Fig. 2(b) at various magnetic fields. The field value is shown on the upper left corner of each image. The measurements were done at 6 K. (m) Schematic of a typical fir tree domain pattern. Black arrows denote the slightly canted magnetization under a small magnetic field. (n) Schematic of domain pattern seen in (h). (o) Cross-sectional view of the V-lines domain pattern of the marked yellow regions in (n). The color scale is 0.8 Hz. The scale bar is 2 $\mu m$.

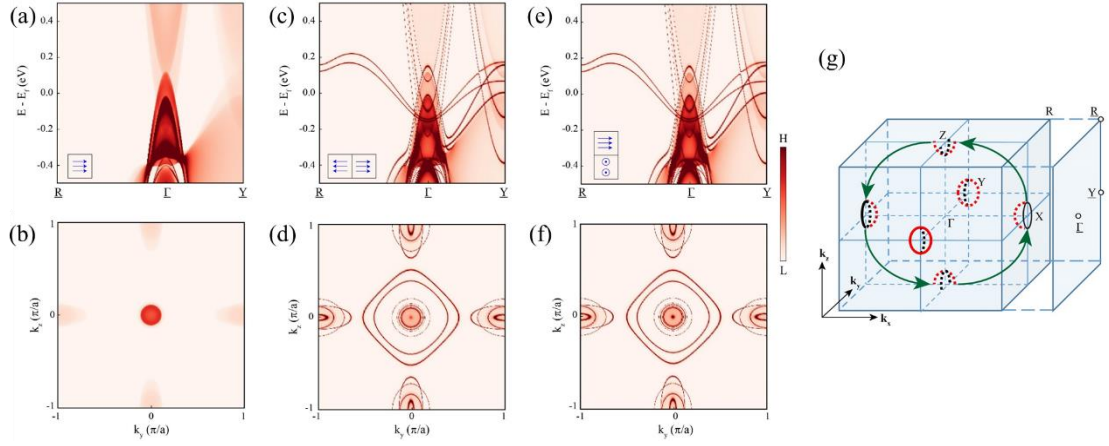

FIG. 4. (a) Surface states for EuB$_6$ with magnetization along the [100] direction. (c), (e) Domain wall states, the magnetization is along the [-100] and [100] and (e) along the [100] and [001] directions for the left (up) and right (down) domain, respectively. (b,d,f) Constant energy mapping corresponding to (a,c,e) taken at the Fermi energy. (g) Schematic of the evolutions of the Weyl nodal loops of EuB$_6$ projected from the FM [100] and FM [001] domain, which are marked by black and red circles, respectively.